\documentclass[preprint]{aastex}

\newcommand{\mg}{\mu G}

\shorttitle{Magnetic tension effect on high-z galaxies}
\shortauthors{Y. Birnboim}

\begin{document}

\title{Magnetically Regulated Gas Accretion in High-Redshift Galactic Disks}

\author{Yuval Birnboim}
\affil{Harvard-Smithsonian Center for Astrophysics, 60 Garden Street,
  Cambridge MA, USA}

\begin{abstract}
Disk galaxies are in hydrostatic equilibrium along their vertical axis. The
pressure allowing for this configuration consists of thermal,
turbulent, magnetic and cosmic ray components. For the Milky Way(MW)
the thermal pressure contributes $\sim 10\%$ of the total pressure near the
plane, with this fraction dropping towards higher altitudes. Out of
the rest, magnetic
fields contribute $\sim 1/3$ of the pressure to distances
of $\sim 3{\rm kpc}$ above the disk plane.
In this letter we attempt to extrapolate these local values to high
redshift, rapidly accreting, rapidly star forming disk galaxies and
study the effect of the extra pressure sources on the accretion of
gas onto the galaxies. In particular, magnetic field tension may
convert a smooth cold-flow accretion to clumpy, irregular star
formation regions and rates. The infalling gas accumulates on the edge
of the magnetic fields, supported by magnetic tension. When the mass of the infalling gas
exceeds some threshold mass, its gravitational force cannot be
balanced by magnetic tension anymore, and it falls toward the disk's
plane, rapidly making stars.
Simplified estimations of this threshold mass are consistent with
clumpy star formation observed in SINS, UDF, GOODS and GEMS surveys.
We discuss the shortcomings of pure hydrodynamic codes in simulating
the accretion of cold flows into galaxies, and emphasize the need for
magneto-hydrodynamic simulations. 
\end{abstract}

\keywords{galaxies: high-redshift --- galaxies: ISM --- galaxies: magnetic fields}

\section{Introduction}

This letter deals with the accretion of the cold flows
into disk galaxies at $z=2-4$. These star forming galaxies are expected to have large
magnetic field, to a degree that will significantly affect the
interaction of the cold flows with the disk. In particular, magnetic
tension will cause smooth cold flows to enter the disk in a clumpy
manner. We derive analytically the typical masses of these clumps and
find that for reasonable parameters of the magnetic fields and the disks,
the clump masses are $10^7-10^9M_\odot$ as observed. The introduction
also contains observational results regarding magnetic
fields in the Milky Way (MW) and nearby galaxies (\S~1.1), and clump clusters and clumpy disk galaxies at
redshift $z=2-4$ (\S~1.2). In \S~2 we derive a toy model for 
the magnetic tension of a disk (\S~2.1), typical parameters of filamentry
cold flows (\S~2.2), and the predicted clump masses (\S~2.3).
The importance of magnetic
tension for gas accretion calculations and the conclusions are found
in \S~3.

\subsection{The Pressure of the Milky Way and Nearby Galaxies}
The gaseous disk of the MW is highly multiphased and
contains molecular gas, neutral gas and ionized gas at various
temperatures and densities. Since these components are observed in
many wavelengths, and by many instruments, it is difficult to combine
the observation into one coherent picture of the interstellar medium
(ISM). The vertical structure of the Galaxy is constructed of neutral
gas and molecules dominating near the disk's plane, to
neutral gas and ionized gas dominating at vertical heights larger than
$z \sim 1{\rm kpc}$, with a considerable ionized gas density extending out to
$\sim 3{\rm kpc}$. A
compilation of the various components is available in
\markcite{ferriere01}{Ferri{\`e}re} (2001). Using this information, one can compare the
vertical gravitational attraction indicated by this column mass as
well as stars and dark matter to
the force arising from the thermal pressure of the combined gaseous
components. It is striking that the thermal
pressure at the mid-plane only supports $\sim 10\%$ of the pressure
required to
support the vertical extent of the disk \markcite{badhwar77, dehnen98,cox_don05}({Badhwar} \& {Stephens} 1977; {Dehnen} \& {Binney} 1998; {Cox} 2005). This fraction at mid-plane
decreases further at larger radii. The remaining $90-100\%$ of the
pressure at each altitude arises from a non-thermal pressure
sources: turbulent, magnetic and cosmic ray
pressures. The relative importance of the three
components is comparable, indicating that they are roughly in
equipartition \markcite{boulares90,beck96}({Boulares} \& {Cox} 1990; {Beck} {et~al.} 1996, and reference within.). This
is currently 
a widely accepted hypothesis,
which is motivated by the physical expectation that supernova
explosions drives turbulence \markcite{mckee77}({McKee} \& {Ostriker} 1977) which enhances seeded
magnetic fields. The magnetic fields act to 
confine the cosmic rays and effectively set the diffusion rate
of cosmic rays out of galaxies.

Magnetic fields are measured by observing Zeeman splitting of $21 {\rm
  cm}$, Faraday rotation of linearly polarised radio emitters
and radio synchrotron emission of cosmic rays reacting to the magnetic fields.
Radio loud nearby disk
galaxies such as  M51, M83 and NGC6946, have total field strengths of
$20-30\mg$ in their spiral arms \markcite{beck09}({Beck} 2009,  and reference within.)
 \markcite{bell03}{Bell} (2003) showed that
the radio emission, most of which is non-thermal synchrotron
radiation, which arises from a rough equilibrium between cosmic rays and
magnetic fields, correlates with SFR over 5 orders of magnitude
in radio emission. The relation $B\sim {\rm SFR}^{0.5}$ is expected, and
indeed observed, for star-forming galaxies \markcite{lisenfeld96}({Lisenfeld}, {Voelk}, \&  {Xu} 1996). We do
not attempt to
model this relation explicitly for our high-z star-forming ($\gtrsim
100{\rm M_\odot~yr^{-1}}$) galaxies,
but adopt, as a lower bound, magnetic field amplitudes of
$20-100\mg$ as reasonable, conservative, extrapolation.
A good review of the measurements and typical values of magnetic
fields can be found in \markcite{beck09}{Beck} (2009).

The magnetic field of the MW, a low SFR, radio quiet
galaxy, ranges from $6\mg$ at the solar
neighborhood, to $10\mg$ in the inner Galaxy. At the spirals near the
center, the magnetic fields can reach $100\mg.$ While the large scale
structure of 
the MW's magnetic fields is hard to measure, the Galaxy
provides a good probe of small scale magnetic fields. Magnetic
fields seem to be entangled between the multiphase components of the
ISM on parsec scales. $60\%$ of the field's strength
tends to be in coherent, ordered magnetic structures, correlating on
kpc scales with the spiral structures of the MW. 
At higher altitudes the magnetic field drops, and at $3~{\rm kpc}$ it is a
factor of $2$ lower when fitting synchrotron radiation, and five times lower
assuming equipartition \markcite{cox_don05}({Cox} 2005).

It is worthwhile to emphasize the physical situation that arises from
these conditions: the magnetic fields are pressurized but do not
interact gravitationally. The field would disperse to infinity unless
there is a small amount of gas locked between the entangled
magnetic field, pulling
it towards the disk. While the gas does not contribute a significant
amount of thermal pressure, it contributes all the gravitational force,
pulling everything together.

\subsection{Clumpy Star Formation at High Redshift}
During the past few years data have begun to accumulate regarding the
structure of high redshift spiral galaxies. Their morphology was
investigated using the Hubble Deep Field survey
\markcite{cowie95}{Cowie}, {Hu}, \& {Songaila} (1995) and the Ultra Deep Field
\markcite{elmegreen05,elmegreen07}({Elmegreen} {et~al.} 2005, 2007). They showed that the
majority of high-z galaxies are disks, with clumpy, irregular
morphology (``Clump clusters'' or ``Chain galaxies'') and high
velocity dispersions. Most of the star formation
is observed to be concentrated in a few, well defined
clumps with spatial extent of $\sim 1{\rm kpc}$ and mass
of $10^7-10^9M_\odot.$ The inferred velocity
dispersion of these galaxies is $20-50 {\rm km~sec^{-1}}.$
These observation
are complemented spectroscopically by observations by the SINS survey
\markcite{genzel06,forster06}({Genzel} {et~al.} 2006; {F{\"o}rster Schreiber} {et~al.} 2006) that contains 80 $H_\alpha$ emitting galaxies with
masses $M\sim 3\times 10^{10}M_\odot$, most exhibiting rotating
irregular gas patterns \markcite{shapiro08}({Shapiro} {et~al.} 2008).
 Lower redshift information, derived from the
GOODS and GEMS surveys, is analyzed in \markcite{elmegreen09}{Elmegreen} {et~al.} (2009) and from
the MASSIV survey in \markcite{epinat09}{Epinat} {et~al.} (2009).
\markcite{elmegreen09}{Elmegreen} {et~al.} (2009) demonstrated that there is an evolutionary
sequence of galaxies with the highly irregular clumpy ones at one end
(making $50\%$ of the galaxies at $z\sim 4$),
and a smoother disk at the other (making $90\%$ of the galaxies at
$z\sim 1$).  

While these galaxies are star-bursting, with typical SFR of $100-200{\rm M_\odot~yr^{-1}},$ most show no signatures of
galaxy mergers and appear to maintain their disk. Their gas supply is
most likely a result of gaseous cold flows accreting efficiently onto the
galaxies \markcite{dekel09,bournaud09}({Dekel} {et~al.} 2009b; {Bournaud} \& {Elmegreen} 2009). Gas that is accreted onto the halo is unable to sustain a
virial shock because the ``would-be'' post-shock gas would be cooling
faster than it is contracting - unable to provide pressure support for
the virial
shock. This has been shown analytically by \markcite{bd03,db06}{Birnboim} \& {Dekel} (2003); {Dekel} \& {Birnboim} (2006) and
numerically by
\markcite{keres05,ocvirk08,keres09,brooks09,agertz09}{Kere{\v s}} {et~al.} (2005); {Ocvirk}, {Pichon}, \& {Teyssier} (2008); {Kere{\v s}} {et~al.} (2009); {Brooks} {et~al.} (2009); {Agertz}, {Teyssier}, \& {Moore} (2009). There is a broad 
transition \markcite{db06,keres09}({Dekel} \& {Birnboim} 2006; {Kere{\v s}} {et~al.} 2009) between cold-dominated halos and
hot-dominated halos at 
which the accretion is in narrow, cold filaments penetrating through a
hot, diffuse halo.
The cold flows ($\sim 10^4K$
gas) flow undisturbed, preferentially perpendicular to the disk's
plane \markcite{dekel09}({Dekel} {et~al.} 2009b), until the vicinity of the disk. These filaments then
deposit their kinetic energy near the disk\footnote{The slowdown can be
  gradual, by a series of weak shocks (\markcite{keres09,agertz09}{Kere{\v s}} {et~al.} (2009); {Agertz} {et~al.} (2009) using SPH
  and AMR simulations) or abrupt in a strong, ``isothermal'' shock
(\markcite{dekel09,ocvirk08}{Dekel} {et~al.} (2009b); {Ocvirk} {et~al.} (2008), AMR simulations)}, join
the disk and efficiently form stars. 
 Based on lambda cold dark matter ($\Lambda CDM$)
hierarchical formation models \markcite{neistein08,dekel09}{Neistein} \& {Dekel} (2008); {Dekel} {et~al.} (2009b) find that 
 $2/3$ of the accreted gas are in the form of smooth, cold, inflowing filaments.

Recent theoretical work attempts to show how such clumps can be
created. By using hydrodynamic simulations of disks, \markcite{bournaud07}{Bournaud}, {Elmegreen}, \&  {Elmegreen} (2007) demonstrated that gas rich
disks corresponding to these high-z galaxies break rapidly into clumps
via gravitational instability, and that these clumps migrate towards the
center, creating a bulge. \markcite{dekel09b}{Dekel}, {Sari}, \&  {Ceverino} (2009a) showed that when there is a
constant supply of smoothly accreted gas, the disk is maintained
marginally Toomre stable, allowing for a prolonged evolution of bulge
and clumpy disk consistent with observations. Using high resolution
AMR simulations, \markcite{agertz09}{Agertz} {et~al.} (2009) found that clumps naturally form
within the
disk from cold flows, and that the cold flows do not exhibit strong shocks at the edge
of the disks.

\section{Disk Accretion in the presence of Magnetic Tension}
\subsection{Magnetic Tension}
\begin{figure}
\plotone{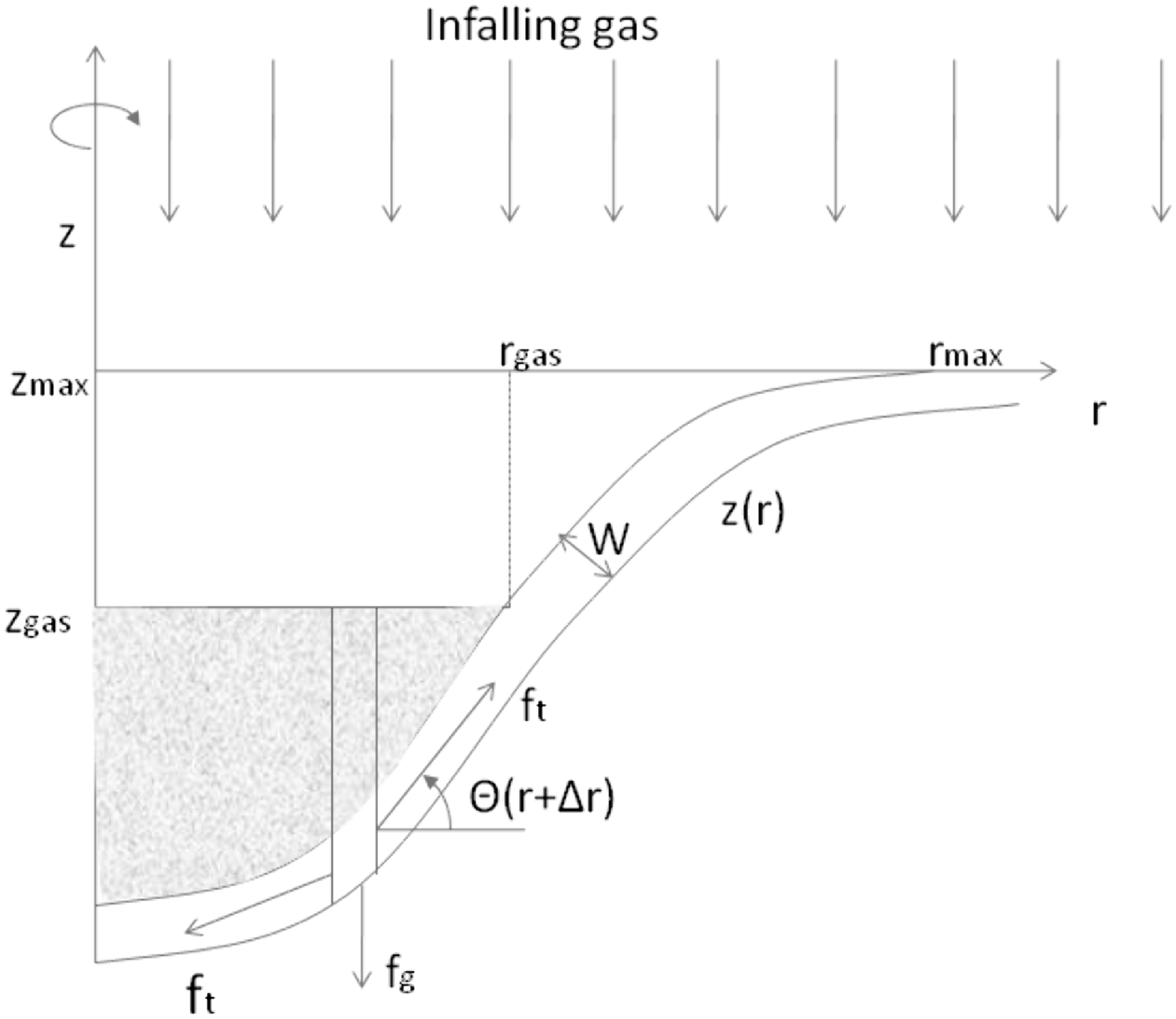}
\caption{A schematic representation of the magnetic tension sheet and
  the forces acting on it.\label{fig:scheme}}
\end{figure}

In this section we estimate the dynamic effect of magnetic tension on
the accretion efficiency of cold-filament gas onto the
galaxies. If one pulls a magnetic tube along its field lines, the
amplitude of the magentic field does not change, while the volume it
occupies grows. The work required to extend the tube by a distance $dl$
is:
$\frac{B^2}{8\pi}Adl$ with $A$ being the surface crossection of the
tube. The tension of the magnetic field is thus:
\begin{equation}
\sigma=\frac{B^2}{8\pi}.
\end{equation}
The magnetic fields are entangled on some scale, which creates a net
of magnetic fields woven through the gas. While the field adjusts
quickly to pressure variations, it does react against being bent. The 
toy model proposed here treats the field as a sheet of elastic matter,
being bent inwards by gas that is accreted onto it. As gas falls on
top of the magnetic sheet, it will
flow and settle at the lowest altitude it can, to minimize its
potential energy. We assume that the
deflection of the sheet is cylindrically symmetric around the center
of the perturbation, and that the magnetic fields are
pinned to their initial altitude at some radius by the joint forces of
the entangled
magnetic fields. We denote the radial direction by $r$ and the
vertical dimension by $z.$ The magnetic fields are pinned to $z_{\rm max}$
at $r_{\rm max}$ (fig. \ref{fig:scheme}). We solve for the shape of the line
$z(r)$ for which the system is static in the presence of mass
$M_{\rm cl}$ of gas with density $\rho_{\rm gas}.$ 
The gas will fill the well in the magnetic fields from the lowest
$z$ possible to some height $z_{\rm gas}$ according to:
\begin{equation}
M_{\rm cl}=\int_0^{r(z_{\rm gas})}2\pi r \rho_{\rm gas}(z_{\rm gas}-z(r))dr.
\label{eq:mass}
\end{equation}
The force on some ring element of gas between $[r:r+\Delta r]$ is the
sum of the gravitation on that element, and the difference between the
vertical component of the tension at $r+\Delta r$ and $r$:
\begin{eqnarray}
&2\pi r (z_{\rm gas}-z(r))\rho_{\rm gas}g(z)~dr=\label{eq:force}\\
&2\pi (r+\Delta r)W \sigma
{\rm sin}[\theta(r+\Delta r)]-2\pi r W \sigma {\rm sin}[\theta(r)],\nonumber
\end{eqnarray}
with $g(z)$ being the gravitational acceleration caused by the disk\footnote{throughout this work we use a fit for the
  vertical gravitational profile from \markcite{cox_don05}{Cox} (2005) that takes
  into account the gas components, as well as the star and dark
  matter contribution. For different disk column
  densities the gravitational force is scaled, but its shape and z-dependence
  is not changed. The observed values of column densities by
  \markcite{elmegreen09}{Elmegreen} {et~al.} (2009) of $z\sim 2-4$ galaxies ($\sim 100{\rm M_\odot
    pc^{-2}}$) are roughly the same as those of the MW. Since $\sigma$
  and $g$ in eq. \ref{eq:force} are proportional, a change in
  the scaling of the gravitation corresponds to a change in the square of the magnetic fields.}, $\theta(r)$ the angle between the direction of the line and the
horizontal direction, and $W$ the width of the elastic layer (see
fig. \ref{fig:scheme}). Throughout this work we assume that the width
of the magnetic layer being bent scales like the amplitude of the
perturbation, up to a factor $l_W$:
\begin{equation}
W=l_W\times (z_{\rm max}-z(0)).
\label{eq:width}
\end{equation}

 A more sophisticated decision
would require observational data or MHD calculations about the shape
of the magnetic fields.
The solution of eq. \ref{eq:mass} and \ref{eq:force} is found by
integrating from $0$ to $r_{\rm max}$ with $[z_{\rm gas},z(0)]$ as
boundary conditions, and fitting them to find the required values of
$[M_{\rm cl},z(r_{\rm max})]$. by the simplex numerical method \markcite{NR90}({Press} {et~al.} 1996). 

\subsection{Magnetically Supported Gaseous Blobs}
In an effort to relate this magnetic tension to realistic
cosmological conditions some typical parameters are chosen.
The following is a rather lengthy ``back of the envelope'' calculation
intended to estimate the density of the infalling gas as it interacts
with the magnetic field. The puncture mass, actually, does not strongly
depend on this density, but it is of deductive value to work out these
numbers non-the-less.
Based on simulations we assume 3 conical filaments, each
at roughly $2\Theta=20^\circ$ in diameter. The temperature of the filaments is
$T_{\rm fil}=2\times 10^4K$ and they are in hydrostatic equilibrium with the virialized
hot gas which is at $T_{\rm vir}$. The density of the filaments in these
conditions is
$\rho_{\rm fil}=f_b\rho_u\delta_{OD}[(T_{\rm fil}/T_{\rm
  vir})(1-3\pi\Theta^2/4\pi)+3\Theta^2/4]^{-1},$ with $f_b,\rho_u$ and
$\delta_{\rm OD}$ being the universal baryonic fraction, density and virial
overdensity respectively.
We assume that the velocity of the filament is roughly the sound speed
of the hot component ($c_s^2=\gamma k_BT_{\rm vir}N_A/\mu$). The global
accretion rate $\dot{M}$ for halo mass $M_{\rm vir}$ at redshift $z$ is taken from
\markcite{neistein08,dekel09}{Neistein} \& {Dekel} (2008); {Dekel} {et~al.} (2009b) which sets the radius of each of the three
filaments near the disk according to:
$\dot{M}/3=\pi R_{\rm fil}^2c_s\rho_{\rm fil}$. 
Finally, the infalling gas in the filament is stopped, converting its
kinetic energy to thermal energy. The strong shock or set of weak
shocks would adjust themselves so the ram-pressure of the infalling
gas is balanced by the dense, halted gas below. Assuming that the
post-shock pressure is all thermal the filament
density at the edge of the disk is $\rho_{edge}={\rho_{fil} c_s^2}/(k_B T_{\rm fil} N_A~\mu^{-1}).$
For
$M_{\rm halo}=10^{12}M_\odot,z=2$ we have: $\rho_{edge}=8\times 10^{-24}{\rm
  gr~cm^{-3}}, c_s=180{\rm km~sec^{-1}}, \dot{M}=78{\rm
  M_\odot~yr^{-1}}, R_{\rm fil}=7{\rm kpc}$ and for
$z=4$ we have: $\rho_{edge}=7\times 10^{-23}{\rm
  gr~cm^{-3}}, c_s=230{\rm km~sec^{-1}},  \dot{M}=250{\rm
  M_\odot~yr^{-1}}$ and $R_{\rm fil}=4{\rm kpc}.$ 

The final densities at the edge of the disks are upper limits. If some
of the pressure of this gas is in turbulent form, or if the energy
carried by the filament is spread over a larger volume, this density
would be lower. 
It can be easily shown that the gas at the edge of the disk is in
rough pressure equilibrium with the total pressure of the disk as is
the case for the MW. However, while the MW's
density is smooth and monotonically decreasing as z increases, no
monotonic density profile is possible for cold accretion. When the
temperature above the 
disk is similar to that of the disk, the extra pressure terms at the
disk require that the density above the disk be larger than at the
disk. This gas would sink in (via Rayleigh Taylor instabilities)
unless the disk has surface tension such as the magnetic tension
discussed here.
The ionization fraction of the infalling gas is also of interest
because the magnetic fields interact only with ionized gas. We expect
that gas at $1-2\times 10^4K$ and densities as mentioned above the gas
will always be
sufficiently ionized so that ambipolar diffusion will be
ineffective\footnote{The timescale for ambipolar diffusion of neutral gas with
  number density and ionization fraction of $n=1~{\rm cm^{-3}},~x=10^{-3}$ 
  through magnetic fields $B\sim 20\mu G$ over a distance of $\sim 1~{\rm kpc}$
  is $t>10^{11}~{\rm yr}$ \markcite{brandenburg94}({Brandenburg} \& {Zweibel} 1994).}.

\subsection{Typical Masses of Magnetic Tension Induced Clumps}
\begin{figure}
\plotone{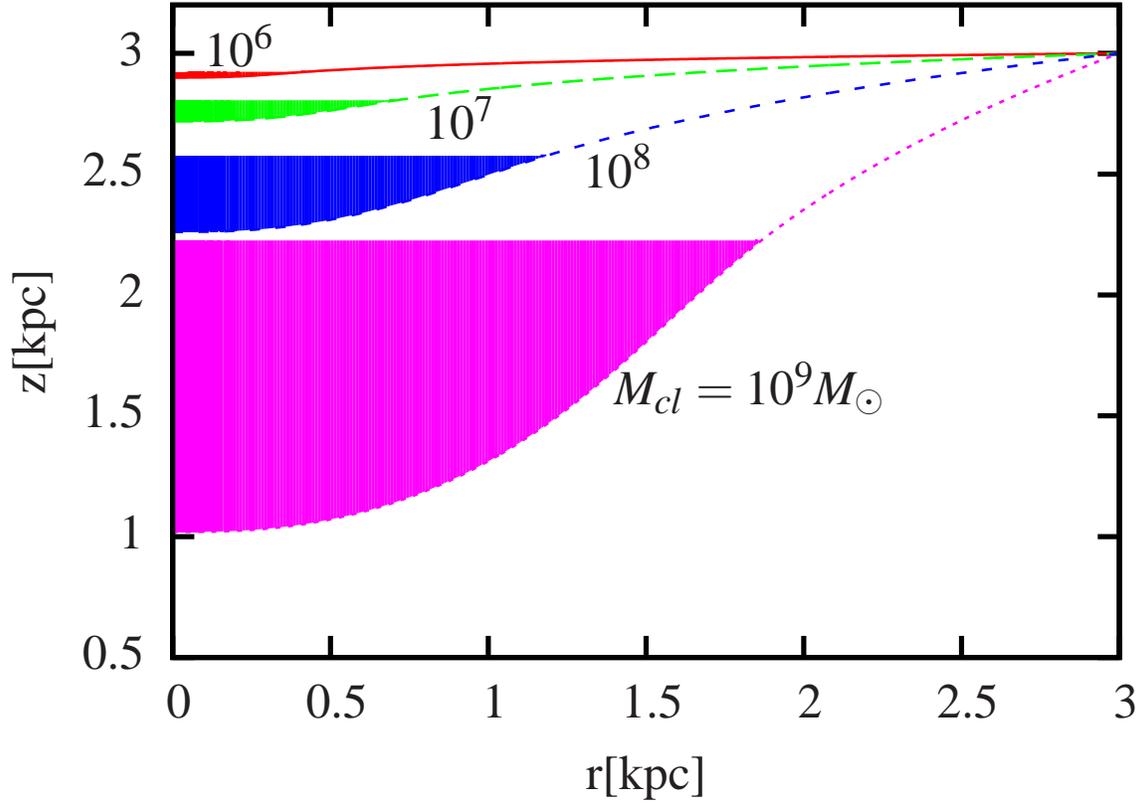}
\caption{The evolution of the magnetic layer as the mass is increased
  from $10^6-10^9M_\odot$. The magnetic fields react to the weight of
  the gas by bending inwards, until, around $10^9M_\odot$ the tension
  cannot support the weight any more and a static solution becomes
  impossible. The gas density, magnetic field, shear layer width and
  maximal radius of perturbation are $\rho_{\rm gas}=10^{-23}{\rm
    gr~cm^{-3}}, ~B=100\mg,~l_W=0.3,~R_{\rm max}=3{\rm kpc}.$\label{fig:sink}}
\end{figure}

The mass of clump for which the clump can puncture through the
magnetic tension depends on the magnetic strength (eq.~\ref{eq:force}), width of the
sheared layer (eq.~\ref{eq:width}; $l_W=0.3.$ throughout this section) , the gravitational profile of the
disk, the density (through eq.~\ref{eq:mass} and
\ref{eq:force}) and on the radius at which magnetic fields are assumed
to be pinned to their location by the global forces of the disk
($R_{\rm max}=3{\rm kpc}$ throughout this section). 
Figure~\ref{fig:sink} shows four clump masses at density
$10^{-23}{\rm gr~cm^{-3}}$ and magnetic field of $100\mg.$ As the mass
increases, the gravitation bends the magnetic fields and gas fills the
lowest potential energy possible. The
puncture mass is just above $M_{\rm cl}=10^9M_\odot.$

\begin{figure}
\plotone{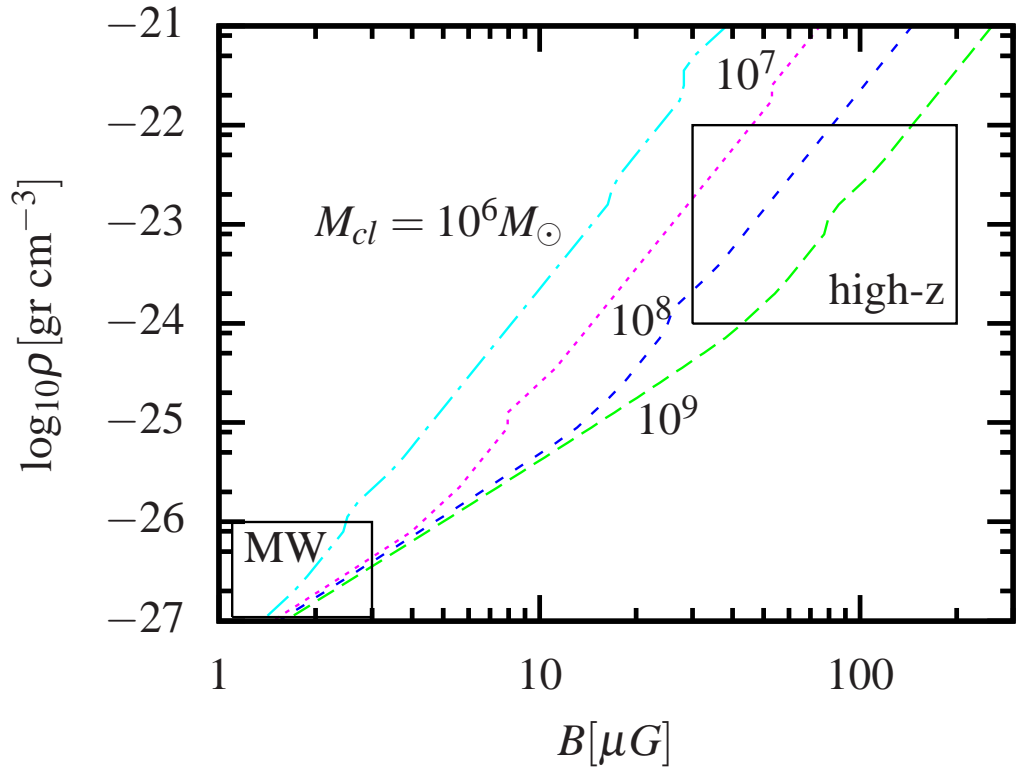}
\caption{The dependence of the puncture mass (various lines) on the
  magnetic field (x-axis) and density (y-axis). The rectangles mark
  typical values for the MW and high-z galaxies in that parameter space.\label{fig:bro}}
\end{figure}

Figure~\ref{fig:bro} shows the puncture mass for
various gas densities and magnetic fields. For the MW, the magnetic
tension will not affect the accretion of gas onto the galaxy (the
``MW'' rectangle in the figure). However,
for larger magnetic fields and reasonable gas densities (marked with the
``high-z'' rectangle in fig.~\ref{fig:bro}) the mass of clumps that can
penetrate is determined by the magnetic tension. For a wide range of
parameters, particularly for high magnetic fields and lower accretion
densities, the puncture 
mass increases to $\gtrsim 10^{10}M_\odot.$ This gas will break through
the magnetic fields instantaneously, yielding large gas fractions
which are necessary for the in-situ clump instabilities found in 
\markcite{bournaud09}{Bournaud} \& {Elmegreen} (2009). The degree of coherence of ages of stars in the clumps is a
good test for this scenario.

\section{Discussion}
We have shown that for $z=2-4$ star forming galaxies the magnetic
fields at the edge of the disk are strong enough to alter the behavior of gas
falling onto these disks. The expected magnetic tension is capable of
stopping the cold accretion until
some threshold mass is accumulated, at which point the clump punctures through
the magnetic fields and falls into the disk. The model could either
produce individual clumps, or suppress accretion onto the galaxy for
long periods of times, causing critical, coherent star-bursts on
global galactic scales. Since the estimation of the magnetic field
strengths here is very conservative, the magnetic tension will probably
also affect gas
accretion in more typical star forming galaxies with rates of $\leq
100{\rm M_\odot yr^{-1}}.$
The model depends on
the shape, correlation length and strength of the magnetic fields
which are poorly known by theoretical considerations and
extrapolation from local galaxies. There are no observational
constraints on magnetic fields at these redshifts.

For the MW, and nearby quiescent spirals, the magnetic fields
are not strong enough to significantly affect the gas accretion, and
the SFR will be dominated by the supply rate and other processes. Even
if the magnetic fields were stronger, the time that would take for
significant clumps to accumulate (at $1-4{\rm M_\odot ~yr^{-1}}$)
would be long with respect to the time it takes for magnetic fields to
realign, which should be of the order of the dynamical time of the
disk. This is not the case when the accretion rate is $\gtrsim 100 {\rm M_\odot ~yr^{-1}}.$

Neutral gas can potentially flow through the magnetic fields. The
conditions required for such a process are typical for molecular
clouds with extremely low ionization fractions. The relatively low
densities and high temperatures ($T_{\rm gas}=2\times 10^4K$)
discussed here will render this ambipolar diffusion negligible. 

The toy model described in \S~2 should not be taken at face
value. It should, however, serve as an indication that non-thermal
pressure components will have a significant, non-trivial dynamic
effect on the accretion of gas onto galaxies. Pure hydrodynamic
simulations are missing the effects of magnetic fields and cosmic
rays, which account for $\sim 2/3$ of the pressure in the disk. This
could be compensated in part by a correct choice of the cooling
function or the equation of states of star forming regions
\markcite{springel03}({Springel} \& {Hernquist} 2003). However, no form of gaseous equation of state can
produce stress-strain behavior similar to that of magnetic tension. If this is
important, full MHD codes should be used. Hydrodynamic
simulations will predict correctly the shape and mass of the gaseous halo,
the rate of the infalling gas and the cold filaments. They will also
predict correctly the overall SFR of galaxies,
because it is basically a question of gas supply. They are not
expected to correctly predict the local, or temporal gas
accretion or SFR, when the accretion is faster than
typical mixing times in the disk. Other processes, such
as angular momentum transfer in the disk, will depend on the dynamics
within the disk which are dominated by non-thermal pressure. Shocks
and sound waves in the disk will depend on the
turbulence and Alfv\`en speeds, rather than on the sound speed of the gas.

I thank Donald Cox for spending time and effort to walk me
through the magnetic field calculations, and Shmulik Balberg for
useful discussions and good advice. I thank the referee, Frederic
Bournaud for useful comments.

\bibliography{}

\clearpage

\clearpage

\clearpage

\end{document}